\documentclass[10pt, conference, a4paper]{IEEEtran}
\IEEEoverridecommandlockouts

\usepackage{cite}
\usepackage{amsmath,amssymb,amsfonts}
\usepackage{algorithmic}
\usepackage{graphicx}
\usepackage{textcomp}
\usepackage{xcolor}
\usepackage{upgreek}

\usepackage{cleveref}
\usepackage{float}
\usepackage{booktabs}
\usepackage{multirow}
\usepackage[inline]{enumitem}
\usepackage{siunitx}
\usepackage{subcaption}

\usepackage{flushend}
\usepackage{amsthm,thmtools}

\Crefname{figure}{Fig.}{Fig.}

\declaretheorem[style=definition]{constraint}

\def\BibTeX{{\rm B\kern-.05em{\sc i\kern-.025em b}\kern-.08em
    T\kern-.1667em\lower.7ex\hbox{E}\kern-.125emX}}

\graphicspath{{figures/}}
\makeatletter
\def\input@path{{figures/}}
\makeatother

\begin{document}

\title{Automated Deployment of Real-Time Tasks for Phased Execution on Scratchpad-Based Multicore Platforms
\thanks{This work has received funding from the European Chips Joint Undertaking under Framework Partnership Agreement No. 101194371 (Rigoletto) including national funding from the German Federal Ministry of Research, Technology and Space (BMFTR). The responsibility for the content of this publication lies with the authors.}
}

\author{\IEEEauthorblockN{Konstantin Dudzik\IEEEauthorrefmark{1}, Maximilian Kirschner\IEEEauthorrefmark{1} and J{\"u}rgen Becker\IEEEauthorrefmark{1}\IEEEauthorrefmark{2}}
\IEEEauthorblockA{\IEEEauthorrefmark{1}\textit{FZI Research Center for Information Technology}, Karlsruhe, Germany}
\IEEEauthorblockA{\IEEEauthorrefmark{2}\textit{Karlsruhe Institute of Technology}, Karlsruhe, Germany \\
	\{dudzik, kirschner, juergen.becker\}@fzi.de}
}

\maketitle

\begin{abstract}
The increasing throughput demands in real-time systems and the consolidation of functionality on few, high-performance platforms are driving a shift towards parallel architectures.
A key challenge for multicore real-time systems is the interference from contention for access to shared memory.
Phased execution approaches address this challenge by restricting shared memory access to mutually exclusive memory phases, while computation is limited to memories local to each core.
However, the complexity of managing and scheduling said memory phases hinders the adoption of phased execution in real-time applications.

To address this challenge, we propose a model-based deployment methodology that automates the process of adapting applications to phased execution. 
We present an integrated tool-based deployment process that extends the application through the necessary memory phases and provides a runtime environment to orchestrate execution on the target hardware platform.

Our approach requires minimal changes to the application implementation since the deployment tool directly extracts the required information from the generated application binaries.
We validate our methodology using an avionics case study on a RISC-V multicore system, showcasing its key components and principles of operation.

\end{abstract}

\begin{IEEEkeywords}
Real-Time Systems, Deployment, Multicore Processing, Phased Execution, Logical Execution Time
\end{IEEEkeywords}

\section{Introduction}
The ever-increasing demand for computational performance in embedded real-time systems, compounded by the trend to consolidate functionality on central, high-performance platforms, motivates a shift to multicore architectures.
A central challenge lies in the time-predictability of such systems, as the interference caused by contention for shared resources complicates the timing analysis.
Another issue arises from the loss of isolation boundaries between components of formerly distributed systems, which necessitates strict isolation mechanisms to prevent cascading failures.

One approach to address memory-access-related interference is to separate computation and memory access, as proposed by the predictable execution model (PREM). Apart from designated \textit{memory phases}, execution is limited to memories local to the respective core~\cite{prem}.
These measures enable predictable execution on multicore platforms but come at the cost of orchestrating these memory phases.
Thus, integrating phased execution approaches into real-time applications remains an elaborate process.

We propose a model-based deployment methodology that automates the adaptation of real-time applications for phased execution on multicore platforms. 
To achieve this, we present a tool-based deployment process that extracts and schedules memory phases for each application's tasks, accompanied by a runtime environment to implement their execution.

A key consideration to enable the consolidation of multiple applications is the modularity of our approach. Each task is compiled to a separate binary, thereby avoiding the integration of applications at the source code level. 

In summary, the main contributions of this work are:
\begin{enumerate*}[label=\arabic*)]
	\item a modeling scheme for specifying the application's task structure as well as deployment and scheduling specific annotations;
	\item a runtime environment that facilitates the phased task execution, guaranteeing isolation and deterministic communication between the application's tasks; and
	\item a deployment tool that automates the process of generating the application and runtime binaries, deriving the system schedule and runtime configuration based on the modeled behavior.
\end{enumerate*}

We validate our methodology by applying it to the ROSACE case study~\cite{rosace}. As the target hardware platform, we implemented a RISC-V multicore system with local scratchpad memories (SPM) and a direct memory access (DMA) engine on an FPGA. 

The paper is structured as follows. \Cref{soa} gives an overview of related work. The main components of the proposed deployment methodology are described in \cref{methodology}. \Cref{deployment-process} presents the main steps of the deployment process. Next, \cref{runtime} covers the implementation of the runtime environment and deployment tool, while \cref{case_study} shows the application of our approach to the ROSACE case study. Finally, \cref{outlook} discusses future work and summarizes the results.

\section{Related work}
\label{soa}

The issue of assuring time-predictable execution in multicore systems has received significant attention in recent years~\cite{survey}. One approach is the strict temporal isolation of memory accesses at the hardware level using Time Division Multiplexing (TDM) techniques~\cite{t-crest, interpret}.
Software-based approaches employ execution models that aim to avoid interference by explicitly scheduling access to shared memory. This makes them applicable to a broader range of hardware platforms since they do not rely on specialized memory arbitration schemes.

One such approach is the AER model~\cite{aer}, where tasks may only access shared memory during dedicated acquisition (A) and restitution (R) phases, of which only one may be scheduled at a time. This constitutes a generalization of the PREM model~\cite{prem, mthprem}, initially proposed to enable predictable behavior on commercial-off-the-shelf (COTS) systems, which divides task execution into memory and computation phases.

Several works address the scheduling of tasks under these phased execution models. \cite{lille24} proposes scheduling techniques for tasks sets defined by a directed acyclic graph (DAG) using the AECR execution model, which extends the AER model to include a communication phase, whereas \cite{eth-synergistic} proposes to include system-level scheduling information to optimize the generation of PREM-compatible tasks using a genetic algorithm.

Adaptation of applications to predictable execution models has been the subject of several works. \cite{prem-compiler, prem-compiler2}~propose compiler-based approaches to convert code into PREM-compliant binaries by extracting computation phases from a monolithic application.
This work is orthogonal to these approaches in that we start from an abstracted task model and adapt each task independently, allowing for a more modular deployment.

By integrating PREM into the operating system, ~\cite{prem-os} reduces the complexity of the resulting scheduling and deployment problem.
The main difference to the proposed methodology lies in the modularity of our approach, where each task is separately compiled, such that no integration at the source code level is necessary.
Another differentiating factor is the scheduling scheme. \cite{prem-os} proposes coarse-grain TDM memory arbitration instead of explicitly scheduled, time-triggered memory phases.

\cite{sync-aer, sync-aer2} propose methods to deploy programs written in synchronous dataflow languages to multicore platforms using the AER execution model. One factor limiting the range of applications to which these approaches can be applied is the zero execution time abstraction used by these languages.
Additionally, \cite{sync-aer, sync-aer2} statically allocate tasks to the respective SPM, whereas in our approach, task binaries are loaded at runtime to facilitate the deployment of larger applications.

Our approach's scheduling and communication model is based on the Logical Execution Time paradigm~\cite{giotto}, which abstracts from the actual execution time of a periodic task by defining a logical execution time (LET) equal to its period. All communication between tasks occurs at the boundary of their LETs, resulting in a deterministic dataflow.
LET is therefore well suited to describe analyzable multicore applications~\cite{let-automotive2} and to be applied for phased execution models due to its read-execute-write semantics~\cite{lille24}.

Several works consider contention resulting from communication in LET-based systems~\cite{let-automotive, dma-let, kalray-let}.
However, in~\cite{kalray-let}, tasks are permanently allocated to scratchpad memories, which limits the complexity of applications, whereas~\cite{dma-let, let-automotive} do not consider interference resulting from instruction memory access. 

\cite{deployment-let} presents a method to map automotive applications to a LET-based execution concept, optimizing the response time. While it also targets SPM-based multicore systems, tasks are also statically allocated. 

Our methodology shares the LET paradigm and proposes a similar, model-based deployment flow as \cite{schade}. However, the primary focus of this work lies in the adaptation of applications to phased execution, whereas the authors of \cite{schade} target hypervisor-managed platforms, where each task corresponds to a hypervisor partition. Therefore, the deployment process and the functionality of its components are inherently different, as are the practical challenges.

This work builds on the general concept for a predictable execution platform we presented as work-in-progress in~\cite{wip}. 

\section{System Model}
\label{methodology}

The proposed deployment methodology targets real-time applications composed of periodic, interconnected tasks, covering various applications from domains like the automotive or avionics industry.
The underlying task model is based on the LET paradigm, which abstracts away from the actual task execution time by specifying fixed communication times at the boundary of each task's LET. This abstraction enables efficient reasoning about dataflow in multicore systems since the inter-task communication is decoupled from varying task execution times.
In order to adopt phased execution approaches for such real-time applications, the individual tasks need to be extended by memory phases, during which all data relevant to task execution and communication is transferred between the main memory and the scratchpad memories local to the core the task is executed on. 
In this work, a three-phase task execution approach as proposed by the AER model~\cite{aer} is used. The task's data and inputs are loaded in the acquisition phase preceding the execution phase, while the task's outputs are transferred after the task has completed execution during the restitution phase.

A central aspect of the phased execution model is the handling of instruction data. One approach is to load the instructions of all tasks in the scratchpad memories in an initialization phase, thereby simplifying runtime mechanisms and minimizing the overhead during run-time. However, this limits the size of applications that can be deployed to the platform since all tasks must fit within the scratchpad memories.
In this work, a task's instructions are loaded as part of the acquisition phase. Hence, it is only necessary that each task fits into the instruction scratchpad individually. Additionally, this approach enables a flexible assignment of a task's jobs to different cores.

The proposed deployment methodology builds on a system model, which is composed of three main aspects:
\begin{enumerate}[label=\arabic*)]
	\item a model characterizing the underlying hardware platform;
	\item the task execution model, describing the application structure and inter-task communication;
	\item the schedule model concerning the timing of the application tasks' execution phases.
\end{enumerate} 

The remainder of this section describes these central aspects of the system model and their associated parameters and the integration of the deployment tool based on the system model.

\subsection{Target Hardware Platform}
\label{hardware}

Since the deployment methodology incorporates phased execution as a key aspect, the underlying hardware platform must provide the necessary features to support such execution strategies. This mainly concerns the availability of scratchpad memories for each core in order to enable interference-free concurrent execution.
Furthermore, this work assumes a multicore target platform, where one core is reserved as the management core, orchestrating system execution by managing memory transfers to and from the core's scratchpad memories via a dedicated DMA engine. Without such DMA feature, the transfers can also be performed by the management core itself with reduced performance.
In order to enforce isolation between different applications and their tasks, the target platform must provide the ability to restrict a core's memory access in the form of a memory protection unit (MPU).

\begin{figure}[t]
	\centering\scalebox{.8333}{\input{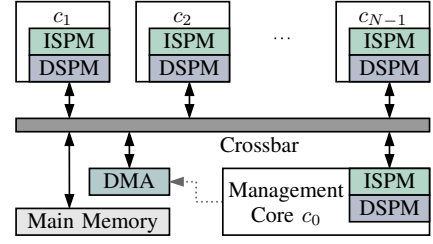}}
	\caption{Exemplary target hardware platform, consisting of \(N\) cores with scratchpad memories for instructions and data, a shared main memory, and a DMA engine.}
	\label{arch}
\end{figure}

\Cref{arch} shows an overview of an exemplary target platform with \(N\) cores, denoted as \(c_0\) to \(c_{N-1}\), connected to a crossbar. One core, \(c_0\), functions as the management core, while the application tasks are executed on the remaining cores, referred to as application cores.

The specification of the target hardware platform consists of the number of cores, the address regions of their scratchpad memories, and the main memory's address region. Additionally, information about the execution time of components of the runtime environment needs to be provided.
Specifically, a function modeling the duration of memory transfers depending on their size needs to be specified. From this, the deployment tool estimates the overhead incurred by each memory phase.
In case a DMA peripheral is available, the developer needs to provide implementations of the necessary driver functions in addition to the timing model.

\subsection{Task execution model}
\label{task-model}

The application is defined in the system model through a set of n tasks \(\Theta = \lbrace \theta_1, \dots, \theta_n\rbrace\), where each task is characterized by its frequency \(f\), its WCET \(\tau\), and a set of ports \(P_\mathrm{\theta}\). The frequency parameter denotes the number of task instances scheduled within one \textit{hyperperiod}, which describes the fundamental period of the application. In the following, a task instance is referred to as a \textit{job}.

Since each task is executed independently from a scratchpad memory, the estimation of the WCET of a task's execution phase does not need to consider any interference. Thus, it can be evaluated for an equivalent system of just one core and its scratchpad memories.

Inter-task communication occurs via ports according to the LET paradigm~\cite{giotto}.
There exist three types of ports: input ports \(P_\mathrm{in}\) and output ports \(P_\mathrm{out}\), which connect one task to another, as well as private ports \(P_\mathrm{priv}\), which are used to preserve state between jobs.
Each port corresponds to a variable in the task's source code. The developer needs to specify its name and port type, while its size is extracted from the task's binary by the deployment tool. For output ports, an optional initialization value can be specified.
\textit{Channels} define the connections between tasks, each connecting one output port to a set of input ports. However, each input port may only be connected to a single output port~\cite{giotto}.

\subsection{Schedule model}
\label{schedule-spec}

The starting point of the schedule model is a periodic schedule, where the LET of each task is equal to its period, given by the hyperperiod divided by the task's frequency parameter.
From this periodic schedule and annotations provided by the developer, a static schedule is derived that describes the system's behavior within one hyperperiod in terms of \textit{events}, during which the scheduler on the management core performs various \textit{operations} relating to memory transfers.

Two operations are scheduled for each job at the boundaries of its LET. At the start of the LET, input ports are populated with the current values of the corresponding output ports. During the second operation at the end of the LET, the task's outputs are published via its output ports.
All operations scheduled for the same logical point in time are mapped to one event. Conceptually, each event takes place instantaneously. In practice, all operations contained in each event are executed sequentially by the runtime, where output ports are published before input ports are processed to comply with LET semantics.
The operations associated with the LET communication mentioned above are referred to as \textit{task input load} operations (\(\omega_\mathrm{TIL}\)) and \textit{task output publish} operations (\(\omega_\mathrm{TOP}\)), respectively.

The introduction of phased execution adds two additional operations for each job, namely the \textit{task load} and \textit{task unload} operations, which correspond to the acquisition and restitution phases of task execution. These operations must fall within the LET of the respective job. Since execution of tasks is non-preemptive, the schedule must be constructed in a way that jobs on the same core do not overlap.

In order to accomplish this, two parameters, \(\varepsilon_\mathrm{TL}\) and \(\varepsilon_\mathrm{TUL}\) are introduced, which allow the developer to specify the time window for task execution within the LET, by delaying the job's acquisition phase and advancing its restitution phase.

\begin{figure}[t]
	\centering\scalebox{.8333}{\input{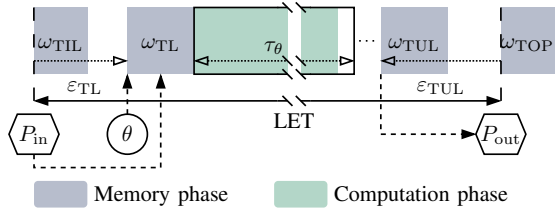}}
	\caption{Overview of memory and computation phases and timing parameters contributing to task execution.}
	\label{task-exec}
\end{figure}

This principle is shown in \cref{task-exec}, which outlines the memory and computation phases involved in executing one job and the effect of the parameters \(\varepsilon_\mathrm{TL}\) and \(\varepsilon_\mathrm{TUL}\).	
Therefore, the schedule is specified by defining the two offsets, \(\varepsilon_\mathrm{TL}\) and \(\varepsilon_\mathrm{TUL}\), as well as the core on which the job is executed, for every job of each task.

Based on these specifications, a list of operations with a given start time is derived, where each operation \(\omega\in\lbrace \omega_\mathrm{TIL},\omega_\mathrm{TL},\omega_\mathrm{TUL},\omega_\mathrm{TOP}\rbrace\) corresponds to one of the four necessary operations to execute one job. The deployment tool then constructs the system schedule, which is composed of a sequence of \(m\) events \(\langle e_1,\dots,e_m \rangle\), where \(m\) corresponds to the number of operations with distinct start times. These events describe the steps necessary to execute one hyperperiod, which the management core performs in a time-triggered fashion.

Each event contains memory phases to transfer data to or from the application core's scratchpads or within the main memory. The management core and all affected application cores are considered occupied for the entire event duration.
In the following, the start time of the \(i\)th event relative to the current hyperperiod is denoted by \(t_i\), while \(\tau_i\) refers to its duration, which is estimated based on the size of the included transfers and a static component for the event itself and each associated operation. The information on the runtime overheads is provided as part of the target platform specification.
The different types of operations are explained in detail below.

\subsubsection*{Task input load}
\(\omega_\mathrm{TIL}\)\\
The task start operation performs the copy-in of the task's input ports to buffers. This operation is always mapped to the event at the beginning of the LET of the corresponding job. For the \(k\)th job of task \(\theta_i\), this is given by:
\[t_\mathrm{TIL} = (k-1)\cdot\tau_\mathrm{hyp}\cdot f_i^{-1}, k\in\lbrace 1,\dots,f_i\rbrace\]

\subsubsection*{Task load}
\(\omega_\mathrm{TL}\)\\
Similarly, the task load operation transfers all of the task's instruction and data sections, input ports, and private ports to the scratchpads of the core to which the corresponding job is mapped.  The start time of this operation is given by:
\[t_\mathrm{TL} = t_\mathrm{TIL} + \varepsilon_\mathrm{TL}, \varepsilon_\mathrm{TL}\ge0\]

\subsubsection*{Task unload}
\(\omega_\mathrm{TUL}\)\\
The task unload operation transfers the output ports of the task to buffers, which hold their values until the end of the LET. Its start time is given by:
\[t_\mathrm{TUL} = t_\mathrm{TOP} - \varepsilon_\mathrm{TUL}, \varepsilon_\mathrm{TUL}\ge0\]

\subsubsection*{Task output publish}
\(\omega_\mathrm{TOP}\)\\
Analogously to the \textit{task input load} operation, a \textit{task output publish} operation copies the buffered task outputs into the task's output ports. It is always scheduled at the end of the LET of the corresponding job. Its start time for the \(k\)th job of task \(\theta_i\) is given by:
\[t_\mathrm{TOP} = k\cdot\tau_\mathrm{hyp}\cdot f_i^{-1}, k\in\lbrace 1,\dots,f_i\rbrace\]

\subsection{Scheduling constraints}
\label{sched_constraints}
Based on the schedule specification provided by the developer, the deployment tool generates a system schedule containing all events. To support the developer in specifying a feasible schedule, the timing of the resulting memory and computation phases is verified based on a set of constraints.
It is, however, the responsibility of the developer to define a schedule that fulfills constraints.

In order to simplify the specification of the schedule, it is possible to assign multiple operations to the same point in time. In that case, the operations are mapped to the same event and executed in the order of \(\omega_\mathrm{TUL}\),\(\omega_\mathrm{TOP}\),\(\omega_\mathrm{TIL}\),\(\omega_\mathrm{TL}\). This guarantees that the inter-task communication matches the dataflow specified by the LET abstraction. Since all cores affected by an event are assumed to be occupied for the entire duration of the event, the order in which operations of the same type are performed is not relevant to the schedule analysis of this section.

Since the model only allows positive values for the offsets \(\varepsilon_\mathrm{TL}\) and \(\varepsilon_\mathrm{TUL}\), job execution is always limited to the time window of its LET.
Furthermore, \(\varepsilon_\mathrm{TL}\) and \(\varepsilon_\mathrm{TUL}\) must be chosen such that the execution phases of two jobs on the same core do not overlap. The start of a computation phase is given by the time of the corresponding task load event \(t_{\mathrm{TL}}\) and logically ends at the time of its task unload event \(t_{\mathrm{TUL}}\). Therefore, a task load operation may not be scheduled to start before the task unload operation of the job previously executed on the same core. 
\begin{constraint}
	Given two consecutive computation phases of two jobs \(j_a\) and \(j_b\) of different tasks that are mapped to the same core, where \(j_b\) is to be scheduled after \(j_a\), then the start time of the task load operation of \(j_b\) must be scheduled at the same time or after the task unload operation of \(j_a\).
	\[t_{\mathrm{TL};a} < t_{\mathrm{TL};b} \implies t_{\mathrm{TL};b} \ge t_{\mathrm{TUL};a}\]
\end{constraint}

If the duration of one event exceeds the scheduled start of the next event, a scheduling delay \(\delta\) is incurred, which is taken into account for the feasibility analysis by the deployment tool.
This delay can affect multiple consecutive events but must not affect events of the subsequent hyperperiod. Therefore, the delay of the first event is always zero. In order to compute the scheduling delay for an event \(e_k, k\ge2\) within the hyperperiod, we first find all preceding, consecutive events whose duration extends beyond the start time of the following event, thereby contributing to the schedule delay. The index \(a\) denotes the first event in this chain.

\begin{align*}
	a = \min\left\{ j \in \lbrace 1, \dots, k - 1 \rbrace \,\middle|\, \left( \forall i \in \lbrace j+1, \dots, k \rbrace, \vphantom{\textstyle\sum_{i=j}^{k}} \right. \right.\\
	\left. \left. t_{e_i} < t_{e_j} + \textstyle\sum_{l=j}^{i-1} \tau_{e_l} \right) \right\}
\end{align*}
If such an index \(a\) exists, the schedule delay for event \(e_k\) becomes:
\[\delta\left(e_k\right) = \left( \textstyle\sum_{a}^{k-1} \tau_{e_i} \right) + t_{e_a} - t_{e_k}\]
Otherwise, the schedule delay is zero. The maximum permissible schedule delay is limited by the requirement that there may only be one pending event at a time.
\begin{constraint}
	The delay for an event must be less than the distance until the next event. 
	\[\delta\left(e_i\right) < t_{e_{i+1}} - t_{e_i}\]
\end{constraint}

It is permissible to specify a schedule with overlapping events, as long as the incurred scheduling delay does not cause the slack of a job to become negative. From the time and duration of the events containing the task load and task unload operations, \(e_\mathrm{TL}\) and \(e_\mathrm{TUL}\), as well as the scheduling delay for those events, the slack for a job \(j\) of task \(\theta\) can be computed as:
\[\sigma_{j,\theta} = \tau_\mathrm{LET} - \tau_\theta - \varepsilon_\mathrm{TL} - \tau_{e;\mathrm{TL}} - \delta\left(e_\mathrm{TL}\right) - \varepsilon_\mathrm{TUL}\]
Here \(\tau_{e;\mathrm{TL}}\) refers to the duration of the event containing the task load operation.

Based on the timing of the events, including possible delays, the deployment tool checks that the slack for every job is greater than zero. 
\begin{constraint}
	With \(J_\theta = \langle j_1, \dots, j_f \rangle\) referring to the sequence of jobs of task \(\theta\) with frequency \(f\), the following condition must be fulfilled for a schedule to be considered valid:
	\[\forall \theta \in \Theta, \forall j \in J_\theta : \sigma_j > 0\]
\end{constraint}

If all constraints are fulfilled, the schedule is considered feasible. The deployment tool then constructs the runtime configuration based on the resulting events.

\section{Deployment Process}
\label{deployment-process}

\begin{figure}[t]
	\centering\scalebox{.8333}{\input{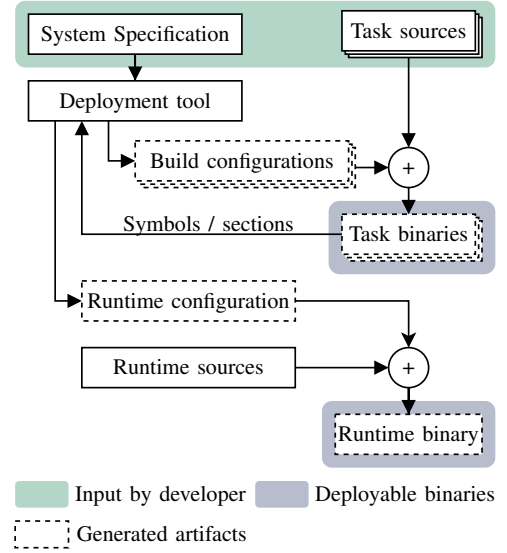}}
	\caption{Overview of the components and steps involved in the deployment process.}
	\label{deployment-flow}
\end{figure}

This section details the process of generating deployable binaries based on the application sources and accompanying system model provided by the developer as described in \cref{methodology}.
\Cref{deployment-flow} shows the central elements of the deployment process, which is split into two main steps, one operating at the task level and the second at the application level.

One key factor to minimize the integration overhead is the level at which the application interfaces with the runtime environment.
Software platforms for real-time applications, like hypervisors and real-time operating systems, typically require the application's tasks to be integrated at source code level.
In contrast, the proposed methodology completely separates runtime and task binaries.
Furthermore, this approach does not require the developer to integrate framework code, therefore placing minimal restrictions on the task's implementation.
Thus, the deployment tool mainly affects the build process of a task, with minimal impact on the task's implementation.

The first step of the deployment process generates the build configuration for each task, consisting of a makefile and linker script.
In its current form, task implementations must be provided as C source code. However, other languages may be incorporated by extending the generation of the build configuration to support the respective toolchain.
The build configuration mainly includes properties of the hardware platform, specifically the SPM regions of the cores on which the tasks are executed.
Each task is then built and linked separately for each core that it is executed on. Since all task binaries are initially located in main memory and only loaded to the respective SPM during run-time, a load address is generated for every task binary, which defines its location in the main memory.

In the second step of the build process, the runtime configuration is generated, which includes the system schedule and the parameters of all memory transfers to execute said schedule.
The addresses and sizes of the sections that need to be transferred to facilitate the phased task execution and the inter-task communication are extracted from each task binary.

Based on the schedule model and the timing annotations provided by the developer, the deplyoment tool constructs the system schedule in the form of a list of events. The timing of said events and the application as a whole is checked against the constraints as described in \cref{sched_constraints}.
Next, a C-file is generated, containing all operations necessary to orchestrate the application's execution on the target platform according to the behavior specification.
This configuration is then included in the build process of the runtime binary.

In addition to the application's tasks, the deployment tool builds a minimal kernel for each application core, which functions as an idle task. 
Finally, the resulting binaries, including each task, the runtime, and the task kernels, can be loaded into main memory according to their load address.

\section{Runtime Environment}
\label{runtime}

This section discusses the implementation of the runtime environment, which facilitates the execution of the application tasks according to the task model described in \cref{task-model}. Its main task is the execution of the operations contained in the system schedule. Additionally, the runtime implements the LET communication semantics, guaranteeing deterministic dataflow within the system, and provides isolation between tasks through memory protection mechanisms.

In order to perform the memory transfers for each schedule operation, the runtime configuration contains the addresses and size for instruction and data sections as well as ports for each task. 
These parameters are extracted from the task binaries during the deployment process as described in \cref{deployment-process}. 
Each port needs to be implemented as a global variable to be compatible with the deployment tool. The type of these variables is irrelevant to the runtime, as each port is treated as an array of bytes whose size is extracted from the symbol table of the task's binary.
A special \textit{input port section} is provided for the variables associated with input and private ports in order to prevent these variables from being zero-initialized during a task's initialization phase.

The LET abstraction requires that the reading of input ports and publishing of output ports occur instantaneously at the boundaries of the task's LET~\cite{giotto}. Since the actual task execution may occur within some interval of its LET, buffers exist for each input and output port to enable this behavior. These buffers, as well as the ports themselves, are located in the data scratchpad of the management core. A TIL operation copies the current value of the associated output port to the buffer of the task's input port; from there, it is transferred to the task during the subsequent LT operation. Analogously, a ULT operation updates the task's output port buffers, which are only transferred to the actual output ports during the TOP operation. Since private ports are exclusive to one task, no buffers are provided.

In order to enable task execution as outlined in \cref{schedule-spec}, the runtime periodically executes the scheduling events as specified in the runtime configuration.
A timer device triggers interrupts on the management core for each event, which then programs the DMA peripheral to execute each memory transfer of the schedule operations mapped to the current event.

The runtime itself is located in the instruction scratchpad of the management core. 
At system startup, a small loader program in main memory is executed on all cores, which loads the runtime binary into the management core's instruction scratchpad and then jumps to the runtime's entry address.

In an initialization phase, the task kernels are loaded into the application cores' instruction scratchpads, which subsequently start executing said kernel.
After the initialization, the runtime starts the application by configuring the timer device for the first event of the schedule and drops to an idle task itself.

The task kernel is executed at an elevated privilege level and waits for the management core to supply the entry address of the task at the end of a task load operation.
Then, execution is transferred to the task, running in an unprivileged mode.
Through the use of memory protection mechanisms, the task may only access its instruction partition and data scratchpad. This guarantees that faults from one task cannot interfere with the rest of the system, thereby addressing the functional safety concerns regarding the consolidation of applications onto one platform.
The implementation of the runtime is based on OpenSBI\footnote{https://github.com/riscv-software-src/opensbi}, using extensions from the keystone~\cite{keystone} framework for memory protection and context switching.

\section{Case study}
\label{case_study}

\begin{figure}[b]
	\centering\scalebox{.75}{\input{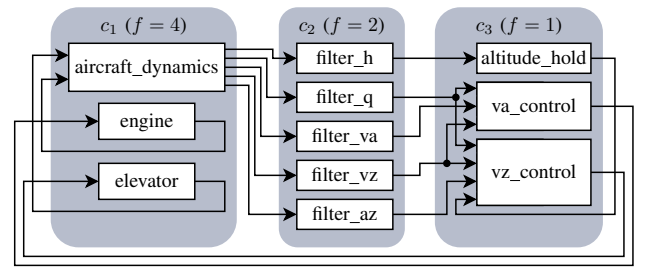}}
	\caption{Overview of ROSACE case study components.}
	\label{rosace_let}
\end{figure}

\begin{figure*}[t]
	\makebox[\textwidth][c]{\hspace{1em}\scalebox{1}{\input{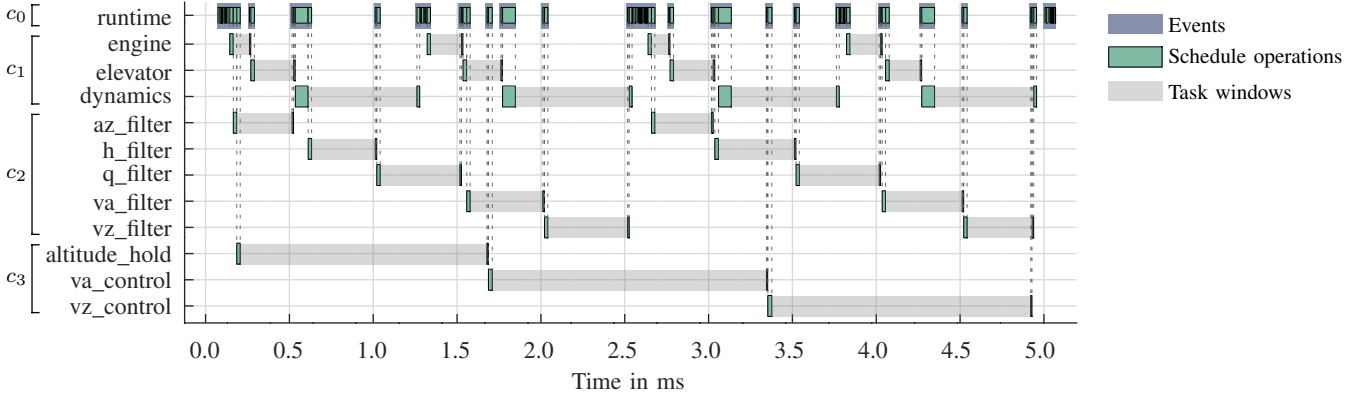}}}
	\caption{Execution trace of one hyperperiod of the ROSACE case study.}
	\label{rosace_exec}
\end{figure*}

The deployment methodology is validated using the ROSACE case study~\cite{rosace, hartstone} as a sample application to showcase the deployment process and runtime mechanisms. \Cref{rosace_let} shows the structure of the ROSACE control application, where the arrows designate the communication channels between the tasks' ports.
For the target hardware platform, a RISC-V multicore system that provides the required features outlined in \cref{hardware} is implemented on an FPGA board (Xilinx VCU 118) using the Chipyard framework. Namely, four Rocket cores~\cite{rocket} running at \qty{100}{\MHz}, configured with tightly integrated scratchpad memories of \qty{120}{\kibi\byte} and \qty{256}{\kibi\byte} in place of level one instruction and data caches respectively, as well as the main DRAM, are connected to a TileLink crossbar. 
A last-level cache has also been omitted to improve predictability.
Additionally, a memory-to-memory DMA peripheral is implemented to improve memory transfer performance.
The resulting system closely resembles the hardware architecture shown in \cref{arch}.
For isolation between tasks, the runtime leverages RISC-V's physical memory protection extension (PMP) provided by the rocket cores.

As the first step of deploying the ROSACE application, a system model is created according to \cref{methodology}. For the hardware platform, this includes the number of cores, the memory map, and the timing characterization of the runtime mechanisms. \Cref{tab:overhead} shows the maximum static overhead observed for a scheduling event and each schedule operation type, not including the transfer times. Each measurement was performed \num{500} times.
Additionally, the duration of memory transfers within the system are estimated. For the case study, the DMA engine was evaluated based on measurements for a set of transfer sizes ranging from \qty{1}{\byte} to \qty{32}{\kibi\byte}, \num{100} times each. The transfer time depending on the size \(c\) is approximated by a linear function with minimal slope, such that it always returns a transfer time larger or equal to the measurements for all data points:
\[\tau_\mathrm{DMA}(c) = \qty{10.1}{\nano\second\per\byte} \cdot c + \qty{2.15}{\micro\second}\]

\begin{table}[b]
	\centering
	\caption{Measured static overheads of runtime mechanisms without memory transfer times}
	\label{tab:overhead}
	\begin{tabular}{lrr}
		\toprule
		Runtime component & Overhead (max.) & Overhead (avg.) \\
		\midrule
		Schedule event & \qty{24.89}{\micro\second} & \qty{11.12}{\micro\second} \\
		Task input load operation & \qty{5.18}{\micro\second} & \qty{4.29}{\micro\second} \\
		Task load operation & \qty{11.69}{\micro\second} & \qty{9.34}{\micro\second} \\
		Task unload operation  & \qty{8.22}{\micro\second} & \qty{6.80}{\micro\second} \\
		Task output publish operation & \qty{4.49}{\micro\second} & \qty{4.20}{\micro\second} \\
		\bottomrule
	\end{tabular}
\end{table}

A model of the application structure, including the frequency, ports, and execution time for each task, as well as the set of channels, is created in the form of a JSON file. In place of an actual WCET analysis, the execution times of each task were measured on the hardware platform.
All tasks with the same frequency in the original task set are mapped to the same application core. Therefore, each task's jobs must be executed sequentially in the same LET window. 

For the filter and control tasks running on cores \(c_2\) and \(c_3\), the offsets for the task load and task unload operations, \(\varepsilon_\mathrm{TL}\) and \(\varepsilon_\mathrm{TUL}\) are chosen to distribute the job's execution windows within the LET equally.
On \(c_1\), more time was allotted for the dynamics task due to its significantly larger execution time. An offset of \qty{100}{\micro\second} is specified for the unload operations of the last job in the hyperperiod of each core in order to reduce the number of operations scheduled at the boundary between hyperperiods. These parameters are shown in \cref{tab:rosace_sched_params} for each task.
From this system model and the sources for each task, the deployment tool generates the task and runtime binaries. The runtime configuration includes the static schedule, which the deployment tool checks for feasibility as described in \cref{sched_constraints}.

\begin{table}[b]
	\centering
	\caption{Case study schedule specification parameters and LET per task}
	\label{tab:rosace_sched_params}
	\begin{tabular}{lrrr}
		\toprule
		Task Name & LET in \unit{\micro\second} & \(\varepsilon_\mathrm{TL}\) in \unit{\micro\second} & \(\varepsilon_\mathrm{TUL}\) in \unit{\micro\second} \\
		\midrule
		engine & \num{1250} & \num{0} & \num{916.67} \\
		elevator & \num{1250} & \num{333.33} & \num{583.33} \\
		dynamics & \num{1250} & \num{666.67} & 0 (100) \\
		az\_filter & \num{2500} & \num{0} & \num{2000} \\
		h\_filter & \num{2500} & \num{500} & \num{1500} \\
		q\_filter & \num{2500} & \num{1000} & \num{1000} \\
		va\_filter & \num{2500} & \num{1500} & \num{500} \\
		vz\_filter & \num{2500} & \num{2000} & \num{0} (\num{100}) \\
		altitude\_hold & \num{5000} & \num{0} & \num{3333.33} \\
		va\_control & \num{5000} & \num{1666.67} & \num{1666.67} \\
		vz\_control & \num{5000} & \num{3333.33} & \num{100}\\
		\bottomrule
	\end{tabular}
\end{table}

\Cref{rosace_exec} shows the execution trace of one hyperperiod of the deployed application. In order to improve the visibility of the scheduling events, the hyperperiod of the application has been changed from \qty{20}{\milli\second} to \qty{5}{\milli\second}. The schedule events are shown at the top for \(c_0\). Each task load and task unload operation is also replicated on the corresponding application core. The time span between these operations, marked as task windows, shows the allotted computation phase of each job on the application cores. 
The remaining TIL and TOP operations are scheduled at the boundaries of the tasks' LETs. Inter-task communication is facilitated during these events by transferring data between connected input and output ports.
The execution graph is generated from timestamps captured from the scheduler's timer peripheral for each scheduling operation.

\section{Conclusion and Outlook}
\label{outlook}
In this work, we presented a deployment methodology for phased execution of real-time applications that automates the adaptation of tasks to a three-phase execution model. The model-based deployment process enables the modular integration of multiple applications since each task is handled independently by the deployment tool. The required information is extracted directly from the task binaries, thereby alleviating the need to integrate tasks and runtime at the source code level.

We validated our methodology on a custom RISC-V multicore platform using the ROSACE case study, showcasing the deployment process.
We plan to extend the runtime environment and system model to support environment ports that interact with external sensors and actors as part of future work.

\bibliographystyle{IEEEtran}
\bibliography{bibi}

\end{document}